\documentclass{optica-article}
\journal{opticajournal}
\articletype{Research Article}
\usepackage{lineno}

\begin{document}

\title{Quantum matched filtering: breaking time-energy separability by 12 orders of magnitude}

\author{Nir Nechushtan\authormark{1}, Hanzhong Zhang\authormark{1}, Yosef London\authormark{1,2}, Mallachi Meller\authormark{1}, Haia Amihai \authormark{1}, Eliahu Cohen\authormark{2$\dagger$}, and Avi Pe'er\authormark{1$\dagger$*}}

\address{\authormark{1}Department of Physics and QUEST Center for quantum science and technology, Bar-Ilan University, Ramat Gan 5290002, Israel}

\address{\authormark{2}Faculty of Electrical Engineering and QUEST Center for quantum science and technology, Bar-Ilan University, Ramat Gan 5290002, Israel}

\email{\authormark{*}avi.peer@biu.ac.il} 
\footnote{\authormark{$\dagger$}Joint advisers of this work} 

\begin{abstract*} 

Detection of signals buried in noise is the major challenge for sensing. Classically, the optimal detector is a matched filter, whose sensitivity meets the classical limit of correlation between the filter target $T(t_2)$ and the measured signal within the noise $S(t_1)$. For classical signals, the correlation is limited by the separability criterion in frequency-time $\Delta(t_1\!-\!t_2)\Delta(\nu_1\!+\!\nu_2)\!\geq\!\frac{1}{2}$. Quantum states, however are not necessarily separable, and the correlation between entangled particles can surpass the classical limits. Specifically, time-energy entangled photons can be simultaneously correlated in time difference $t_1-t_2$ and frequency sum $\nu_1+\nu_2$ with no minimum limit of the above inequality, potentially leading to a drastic enhancement of sensitivity for diversified sensing applications. Yet, to enjoy this quantum enhancement, a unique, \textit{global} detector is needed that can recover the complete information of entanglement in a single shot,  i.e. measure the combined correlated variables $(t_{1}-t_{2})$ {\it and} $(\nu_{1}+\nu_{2})$ \textit{without} measuring the individual frequencies or times. Such a global measurement could, in principle, be realized using the reverse disentangling interaction, such as sum-frequency generation (SFG), but nonlinear interactions at the single-photon level have long been prohibitively inefficient, significantly restricting practical implementations. Here we overcome this barrier: We measure simultaneously and efficiently both the frequency-sum (SFG spectrum) and the time-difference (relative group delay/dispersion) by stimulating the SFG recombination with a strong pump. We generate biphotons with extreme time-energy entanglement (octave-spanning spectrum of 113THz) and measure a relative uncertainty between time-difference and frequency-sum of $\Delta(t_1\!-\!t_2)\Delta(\nu_1\!+\!\nu_2)\!\approx\!2\!\cdot\!10^{-13}$, {\it violating the classical separability bound by more than 12 orders of magnitude}. Our experiment and supporting theory pave the way for improved sensing applications, such as quantum illumination (radar), and for various continuous-variable communication protocols. 

\end{abstract*}

\section{Introduction}

Entanglement of conjugate variables, such as position and momentum, or energy and time, is a fundamental aspect of quantum mechanics. The profound nature of entanglement lies in the ability of quantum systems to exhibit perfect nonlocal correlations between the properties of two or more particles, even when the local values of those properties remain fundamentally undetermined and their simultaneous local measurement is excluded by quantum uncertainty. This paradoxical feature, first recognized by Einstein, Podolsky, and Rosen \cite{PhysRev.47.777}, eventually laid the foundation for our modern understanding of quantum physics, highlighting the inherent nonlocality that characterizes quantum correlations. Today, entanglement is a crucial resource in quantum computing \cite{raussendorf2001one,walther2005experimental,PhysRevLett.97.110501}, communication \cite{bennett1984proc, ekert1991quantum, ursin2007entanglement}, and metrology \cite{giovannetti2004quantum, Giovannetti2011, PhysRevLett.117.110801}, where harnessing these correlations across different variables enables significant advancements in processing and transmission of quantum information \cite{lloyd1996universal, monroe2002quantum}.

The generation of entanglement relies on an entangling interaction between the parties, which for photons inherently entails a nonlinearity, such as three- or four-wave mixing. A well-known example is the generation of entangled photons by spontaneous parametric down-conversion (SPDC) in a three-wave mixing medium, where a pump ($p$) photon from an intense laser splits into a pair of lower-energy photons, named signal ($s$) and idler ($i$) \cite{hong1985theory,boyd2008}. These generated biphotons inherit the properties of the pump, indicating that the sum of all their preserved properties, like energies (frequencies), linear momenta ($k$-vectors), angular momenta (transverse modes) or spins (polarizations) equal those of the original pump, but the properties of each of the photons are not defined and can span a range of possible values. In this paper, we focus on time-energy entangled pairs of photons that are generated by broadband spontaneous parametric down-conversion from a narrowband pump laser: The frequency (energy) sum is fixed ($\nu_s + \nu_i = \nu_p$), as dictated by the pump, but the individual photon frequencies vary across a very wide bandwidth, limited only by the phase-matching of the nonlinear medium, which can span up to an octave of bandwidth ($\Delta\nu\!\approx\!113$THz in our experiment). Conversely, the nonlinear generation of the pair is ``instantaneous'', indicating that the arrival times of the photons are tightly correlated down to the Fourier limit $t_s\!-\! t_i\! \leq \!1\!/\!\Delta\nu$ ($\approx\!9$ fs in our experiment).

While generation of extreme entanglement is rather simple, its unambiguous characterization is a challenge due to the collision between the nonlocal nature of entanglement and the constraints on local measurements due to quantum uncertainty. For example, if we measure locally the frequencies of both photons $\nu_{s},\nu_{i}$ to obtain the energy-correlation, uncertainty precludes any information on the time-correlation. However, simultaneous global measurement of both $\nu_s\!+\!\nu_i$,  $t_s\!-\!t_i$ is allowed by uncertainty if the individual frequencies or times are not measured, thereby uncovering the complete correlation information.  If such a global observation of both time-difference and frequency-sum can be realized efficiently, it will provide a compelling proof of (non-classical) entanglement, since for classical separable states, the mutual uncertainty of time-difference and energy-sum is bounded by \cite{mei2020einstein}
\begin{equation}\label{t-f_entanglement_bound}
  \Delta(\nu_s\!+\!\nu_i)\Delta(t_s\!-\!t_i)\!\geq\!\frac{1}{2},
\end{equation}
whereas quantum mechanically no fundamental limit exists. Indeed, for narrowband biphotons, where the time and frequency correlations could be measured directly with standard detectors, $\Delta(\nu_1\!+\!\nu_2)\Delta(t_1\!-\!t_2)\!=\!0.063$ was previously observed \cite{mei2020einstein}, violating this bound by a factor of $\sim\!8$.  In comparison, broadband biphotons that are generated from a narrowband (single-frequency) pump can dramatically violate the bound of Eq.~\ref{t-f_entanglement_bound}, since their time-correlation can be extremely short ($\sim\!9$fs in our experiment), whereas the observed correlation in frequency will be limited only by the spectral resolution of the sum-frequency measurement. In our experiment, reported hereon, we demonstrate violation of Eq.~\ref{t-f_entanglement_bound} by nearly \textit{$13$ orders of magnitude (!)}. 

Since efficient global detection was not available so far, the standard characterization of entanglement relies on correlations of local measurements. In discrete-variable systems (such as polarization-entangled photons), Bell's inequalities utilize measurements in carefully selected bases to quantify the quantum violation of the classical correlation bounds \cite{bell1964einstein, bennett1993teleporting}. However, the direct extension of Bell's inequalities to continuous-variable (CV) systems remains an ongoing challenge, both theoretically and experimentally. Many important techniques were proposed, including continuous variables (CV) Bell inequalities, e.g. \cite{reid1986violations,cavalcanti2007bell,munro1999optimal,garcia2004proposal,nha2004proposed}, phase-space reconstructions \cite{banaszek1998nonlocality,banaszek1999testing}, variance-based inequalities (i.e. the Simon-Duan criterion and its generalization to higher moments \cite{simon2000peres,duan2000inseparability,shchukin2005inseparability,hillery2006entanglement,agarwal2005inseparability}, and more general Gaussian-state criteria \cite{weedbrook2012gaussian}, but no single method can universally and conclusively detect all forms of (CV) entanglement. Moreover, some of the above methods are challenging in practice, involving complex preparation of sophisticated measurements, and may result in very small Bell violations \cite{brunner2014bell}. In addition, limitations introduced by detector inefficiencies, optical losses, or finite sampling rates can obscure genuine quantum correlations. 

\begin{figure}[ht!]
\centering\includegraphics[width=13cm]{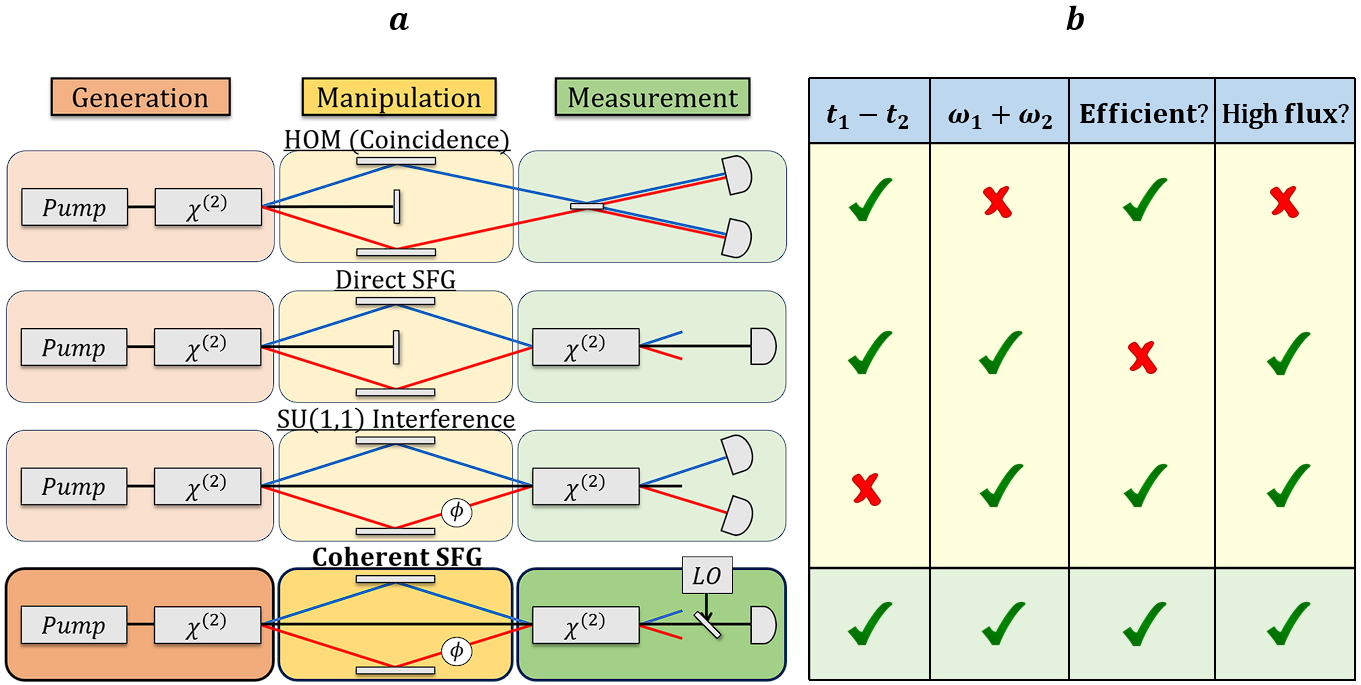}
\caption{{\bf (a) Detection methods of time-energy entanglement:} Coincidence detection with HOM interference, Direct SFG and SU(1,1) Interference are shown in the top three rows, in comparison to the new Coherent SFG (fourth row) that we present now. In all schemes, a pump laser (black line) interacts with a nonlinear medium ($\chi^{(2)}$) to generate entangled photon pairs—signal (blue) and idler (red), followed by manipulation and measurement. The manipulation can includes removal of the pump or phase modulation (or both).  {\bf (b) Pros and cons of each method:} Four critical aspects are addressed -- access to the time correlation $t_1-t_2$,  access to the energy correlation $\omega_1+\omega_2$, detection efficiency, and whether the detectable flux of entangled photons is limited by the speed of the photo-detectors.}
\label{different methods}
\end{figure}

Nevertheless, various techniques were developed to witness and quantify time-energy entanglement, as summarized in Fig.~\ref{different methods}, including coincidence detection in time with Hong-Ou-Mandel (HOM) interference \cite{PhysRevLett.59.2044}, spectral-correlation measurement using SU(1,1) interferometry \cite{shaked2014observing,manceau2017detection} and direct sum-frequency generation (SFG) \cite{dayan2005nonlinear,PhysRevLett.120.053601,liu2020joint} (the table in Fig.~\ref{different methods}b summarizes the advantages and limitations of each method). While these methods have advanced the field, none of them provided the complete correlation efficiently from both time and energy.

Conceptually, a global measurement of time-difference and energy-sum is achievable via time-reversal of the nonlinear entangling interaction. Consider SFG (Fig.~\ref{different methods}a) - the inverse of parametric down conversion (PDC) \cite{dayan2007theory}, where a pair of signal and idler photons is annihilated to form a single SFG photon, to be later detected. Measuring the energy of the  SFG photon provides only the energy-sum of the entangled pair without the individual energies. Conversely, SFG requires both photons to arrive simultaneously, thereby acting as a coincidence detector of $t_{s}\!-\!t_{i}\!\approx\! 0$ which is blind to the specific times. Compared to standard methods of coincidence, HOM or SU(1,1) interference, an \textit{efficient} SFG detector will dramatically enhance the sensitivity and SNR, since it will ``doubly reject'' accidental coincidences in both energy and time. 

\section{Main Contribution}
SFG of broadband PDC forms a powerful optical matched filter for sensing applications: Even classically, the signal and idler fields of broadband PDC are spectral complex conjugates  $E_i(\nu) = E_s^*(-\nu)$, which designates the idler (signal) as a matched filter for the signal (idler). This is particularly advantageous for extracting the signal $E_s(\nu)$ from a noise environment $n(\nu)$ using broadband SFG against the idler, which implements a spectral correlation between the two \cite{Peer2004CDMA}:
\begin{align}\label{Classical_matched_filter_eq}
 E_{SFG}(\nu)\propto& \!\int\! d\nu'\left[ E_s(\nu')\!+\!n(\nu')\right]E_i(\nu-\nu')=\\
= &\!\int\! d\nu'E_s(\nu')E_s^*(\nu'-\nu)+\!\int\! d\nu'n(\nu')E_s^*(\nu'-\nu)=\\
 =&\ S_{corr}(\nu)\!+\!B_{bg}(\nu),
\end{align}
where $S_{corr}(\nu=0)=\int d\nu'\left|E_s(\nu')\right|^2$ is the coherent correlation peak that appears at the frequency of the original pump and is proportional to the signal's total power, which is much larger than the incoherent uncorrelated background $B_{bg}(\nu=0)=\int d\nu'n(\nu')E_s^*(\nu')$ due to the noise. Classically, the noise contribution can therefore be suppressed by the correlation, but cannot be totally eliminated, posing the fundamental limit on the detection sensitivity. 

In contrast, a quantum matched filter can harness entanglement to completely eliminate the uncorrelated background. Specifically, when we perform SFG with entangled biphotons, the coherent correlated peak at the original pump frequency is enhanced by the entanglement (responds linarly), whereas the incoherent background is not (responds quadratically) \cite{dayan2005nonlinear}. Thus, the ratio between the coherent signal and the incoherent noise can diverge in the limit of single entangled biphotons, leading to an enhancement of SNR that is ideally limitless. Indeed, SFG is recognized as ideal for quantum sensing applications that rely on entanglement, such as quantum radar and imaging with quantum illumination \cite{PhysRevLett.118.040801, liu2020joint}. Furthermore, the enhanced SNR of SFG detection directly indicates the time-energy entanglement: While the biphoton amplitude of each individual signal-idler frequency-pair in the spectrum is well below the detection threshold of $1\sqrt{\text{photons}}/\text{mode}$, the coherent sum of all modes across the spectrum can generate a well-detectable SFG signal with practically no noise background due to the ideal rejection of accidental coincidences (as we discuss hereon). Thus, \textit{efficient} SFG with entangled biphotons can transform optical sensing in low-signal environments. 

Unfortunately however, direct SFG at the single-photon level is ridiculously inefficient (typically $\eta_{SFG}\!=\!10^{-9}$ or lower). Consequently, SFG of entangled photons was used so far only to expose the fundamental quantum nature of time-energy entangled photons \cite{PhysRevLett.120.053601}), but not for practical applications of quantum sensing, where a qualitative experimental breakthrough is needed that will enhance the SFG efficiency by many orders of magnitude (ideally to $\eta_{SFG}\!=\!1$). 

Here we overcome this fundamental inefficiency barrier by stimulating the nonlinear interaction in the detection crystal with a strong coherent pump field, which dramatically enhances the SFG efficiency of bi-photons to near-unity \cite{nechushtan2021optimal}. Coherent stimulation—now phase-dependent—underpins nonlinear SU(1,1) interference, where the relative phase between the pump field and biphotons determines the energy flow in the second nonlinear crystal (from biphotons to pump or vice versa). Previously, this non-classical SU(1,1) interference was used for detecting biphotons (Fig.~\ref{different methods}) by measuring the spectrum of the biphotons after the detection crystal. The resulting high-contrast interference fringes reveal the nonlinear pairwise interference of photons, with fringe contrast serving as a non-classical witness \cite{shaked2014observing}. Remarkably, this process remains efficient even in lossy conditions, selectively extracting correlated biphotons. Moreover, SU(1,1) coherence enabled parametric homodyne detection \cite{shaked2018lifting}, which uses the pump as a common phase reference to measure a specified two-mode quadrature of all signal-idler frequency-pairs across a broad spectrum. And yet, standard SU(1,1) detection does not capture the complete quantum correlation, since it still relies on local measurements of the down-converted photons.

To obtain the full correlation information, we return to global detection by SFG, but we stimulate the SFG interaction coherently (see Fig.~\ref{different methods}a). This method leverages the near unit-efficiency of SU(1,1) interference while shifting detection from the biphotons back to the pump—measuring the pump’s depletion, gain, or phase shift due to the coherent SFG conversion. In doing so, it combines SFG’s global, noise-rejecting nature with the efficiency of SU(1,1) interference.

The major experimental challenge of coherent SFG is the need to measure a small SFG gain / depletion on top of an intense pump beam (of $\sim\!400$mW power), which saturates the detector by orders of magnitude. To overcome this we resort to methods of coherent detection: We mix the output beam of pump+SFG on a beam-splitter with a local oscillator at the pump frequency, whose phase is tuned to destructively interfere with the pump, thereby coherently removing most of the strong pump background from the desired SFG result prior to detection. Using coherent seeding of the SFG process and lock-in detection we are able to stabilize the phase of the small SFG signal, as explained hereon. 

 In addition, we present an analytic quantum model (see the Methods section) of coherent SFG that we developed to fit the experimental data. The model provides a closed expression for the SFG amplitude beyond the ``undepleted pump'' approximation \cite{boyd2008} by treating the three-wave mixing process as a perturbative expansion in orders of the spontaneous parametric down conversion (SPDC) amplitude. Our model is applicable in a wide range of parametric gain, from the low-gain regime of single entangled photon pairs ($g\!\ll\!1$) up to the multi-photon regime of high quadrature squeezing ($g\!\geq\!1$). 

\begin{figure}[ht!]
\centering\includegraphics[width=12cm]{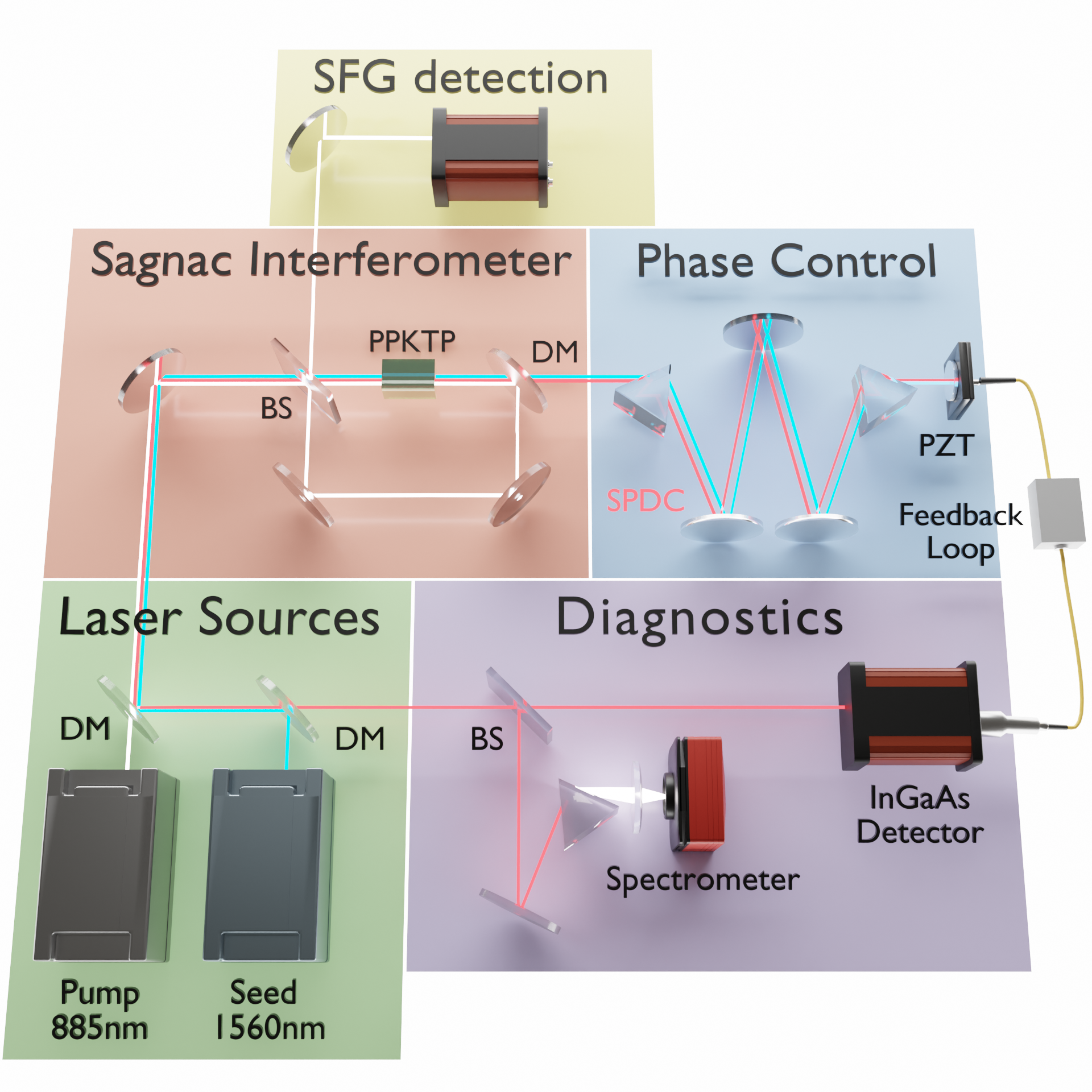}
\caption{{\bf The experimental configuration for Coherent SFG, emphasizing the key elements.} Laser sources (green) include a CW pump laser at 885\,nm (single frequency) and a seed laser at 1560\,nm combined together through a dichroic mirrors (DM1) towards the PPKTP crystal. and the Sagnac Interferometer around it (orange), which enables pumping both forward and backward directions through the crystal. The Sagnac also provides a stable local oscillator to coherently measure the SFG contribution: The Sagnac beam splitter (BS) divides the pump into two paths, each interacting with the nonlinear medium in opposite directions, enabling the generation of entangled photons (forward) and efficient SFG (backwards). The SFG detector is placed at the dark port of the Sagnac interferometer, which prevents detector saturation. Phase Control (blue): A prism-based spectral-shaper and piezoelectric mirror (PZT) ensure precise phase adjustments, compensating for dispersion and maintaining phase coherence across the spectrum. Once the phase is optimized, the entangled photon pairs are reflected back into the PPKTP crystal for a second pass, enabling further nonlinear interaction. Diagnostics (purple): The down-converted spectrum is monitored using a self-assembled spectrometer (prism + CCD camera) and an InGaAs detector at the seed frequency is used to generate feedback for the phase-stabilization of SPDC spectrum relative tothe backwards pump. SFG Detection (yellow): The SFG photons that exit the Sagnac interferometer, measured with a lock-in detection.}
\label{experimental layout}
\end{figure}

\section{Experiment and Results}

Aiming to demonstrate a highly efficient sensor of correlation for both time and energy simultaneously, we are confronted with three major challenges: First, rejection of the high intensity pump that otherwise would overwhelm the detector; Second, precise control of the spectral phase and material dispersion to ensure that the entire spectrum of entangled biphotons contributes constructively to the measured SFG signal; and third, stabilization of the relative phase between the SFG and the pump for a reliable coherent measurement. 

A simplified layout of our experimental configuration is shown in Fig.~\ref{experimental layout}, illustrating the key components and their roles. The configuration is divided into five functional blocks: The laser sources (green) that include the pump ($\lambda$=885\,nm) for SPDC and the seed ($\lambda$=1560\,nm) that is added for stabilization of the SPDC phase relative to the pump (as detailed below);  The nonlinear crystal for PDC and SFG, placed within a Sagnac interferometer (brown) for efficient rejection of the large pump background; This allows the SFG detection (yellow) with lock-in measurement to isolate the SFG contribution only; The diagnostics and phase lock (purple) measure the PDC spectrum at the output of the SU(1,1) interferometer and extracts the phase of the nonlinear interaction in real time for stabilization; And finally, control of the spectral phase of the biphotons (blue) that include a prism-based pulse shaper for precise tuning of the group delay / dispersion between the photons, as well as a fast piezo-actuated mirror for modulation and stabilization of the PDC / SFG phase relative to the pump to allow the lock-in detection of the SFG. The full detection methodology is detailed in the Methods below.

\begin{figure}[ht!]
\centering\includegraphics[width=11cm]{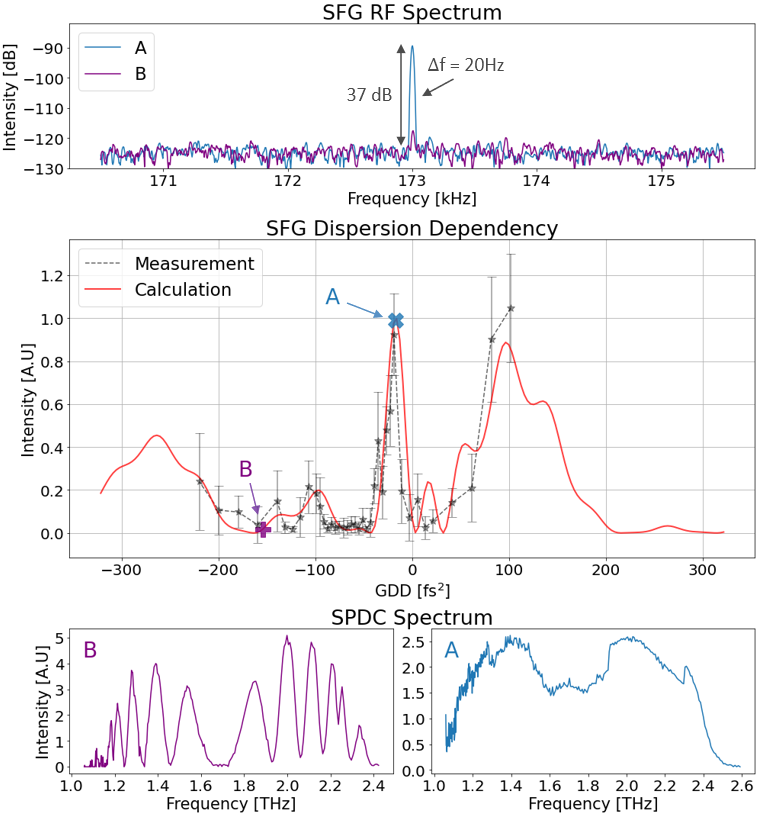}
\caption{{\bf Results. Simultaneous Observation of Energy-Sum and Time-Difference Correlations.} The central panel (time correlation) presents the measured (red points) and calculated (black curve) SFG lock-in signal as a function of the GDD applied to the SPDC spectrum in the spectral shaper. The top panel (energy-correlation) shows the RF spectrum of the lock-in SFG signal, highlighting the sharp peak at the dithering frequency of 173\,kHz with SNR of 37\,dB, which reflects the successful SFG of entangled photon pairs. The blue graph corresponds to the maximum lock-in SFG value near zero GDD (point A in the center panel) whereas the violet graph corresponds to a high GDD value (point B), where the lock-in SFG signal is nearly zero. The bottom panel displays the corresponding SPDC spectra after the second pass through the crystal, showing the spectral SU(1,1) interferogram for the same two GDD values (A- GDD compensated, B- large GDD).}
\label{dispersion_full}
\end{figure}

The main results are shown in Fig.~\ref{dispersion_full}, presenting the correlation in both energy-sum and time-difference (or group delay). The top panel (a) reflects the SFG spectrum of recombination, highlighting the precise match of the sum-energy of the biphotons with the original pump energy and confirming that the process is dominated by the recombination of correlated photons back to the pump frequency  $\nu_{SFG}\! =\!\nu_p$, while the probability of accidental SFG to other SFG frequencies from non-twin photons vanishes (as theoretically predicted, see below). Specifically, panel (a) shows the RF spectrum of the of the detected pump+SFG intensity, demonstrating a single peak of high contrast ($>\!37$dB) at the lock-in dither frequency $\omega$. The measured intensity modulation is due to the beat of interference between the carrier pump at $\nu_P$ and the SFG sidebands at $\nu_{SFG}\!=\!\nu_P\!\pm\!\omega$ due to the phase modulation of the SPDC, thereby capturing the frequency-difference between the SFG signal and the pump at Hz-level resolution ($20$Hz width), limited only by the integration time of the RF analyzer. The background of the SFG peak is purely the electronic noise floor of our digital spectrum analyzer (Moku:Lab). The absence of an optical contribution to the background at other frequencies $\nu\!\neq\!\omega$ indicates that the recombination of correlated biphotons dominates and that accidental SFG events are insignificant.
 
The central panel (b) of Fig.~\ref{dispersion_full} reflects the correlation in time-difference between the photons by observing the dependence of the SFG spectral peak (at the dither frequency $\omega=173$KHz) on the relative group delay between the signal and the idler photons before entering the SFG crystal. We continuously tune the group delay (and GDD) by varying the amount of material that the beam of biphotons traverses in the spectral-shaper (by shifting the Sapphire prisms in/out). The additional material dispersion imprints a quadratic spectral phase $\phi(\nu)\!=\!\beta_2\nu^2$ on the broad spectrum of spontaneous biphotons, which in time indicates a group delay (and dispersion) between each signal frequency and its twin idler.; and the SFG signal extracted from the lock-in detection is simply $I_{SFG}\!\sim\!\text{Re}[\alpha_r^*\sum_{\nu}a_s(\nu)a_i(-\nu)\mathrm{e}^{\mathrm{i}\phi(\nu)}]$, which reflects a homodyne measurement of the SFG sum over all the signal-idler pairs with $\alpha_r$ the coherent amplitude of the pump residue at the Sagnac output (see derivation below). Since even small GDD values alter the phase of the spontaneous SFG field, the variation of the GDD is immediately reflected in the homodyne, causing the sharp peak. The bottom panels (c) and (d) show the spectral SU(1,1) interferograms for two specific GDD values, marked A and B on the central panel (b): A - near GDD compensation (bottom right, blue), where the spectral phase is flat; and B - at a large GDD value (bottom left, violet), where the large quadratic spectral phase is imprinted on the spectral fringes.

\begin{figure}[ht!]
\centering\includegraphics[width=13cm]{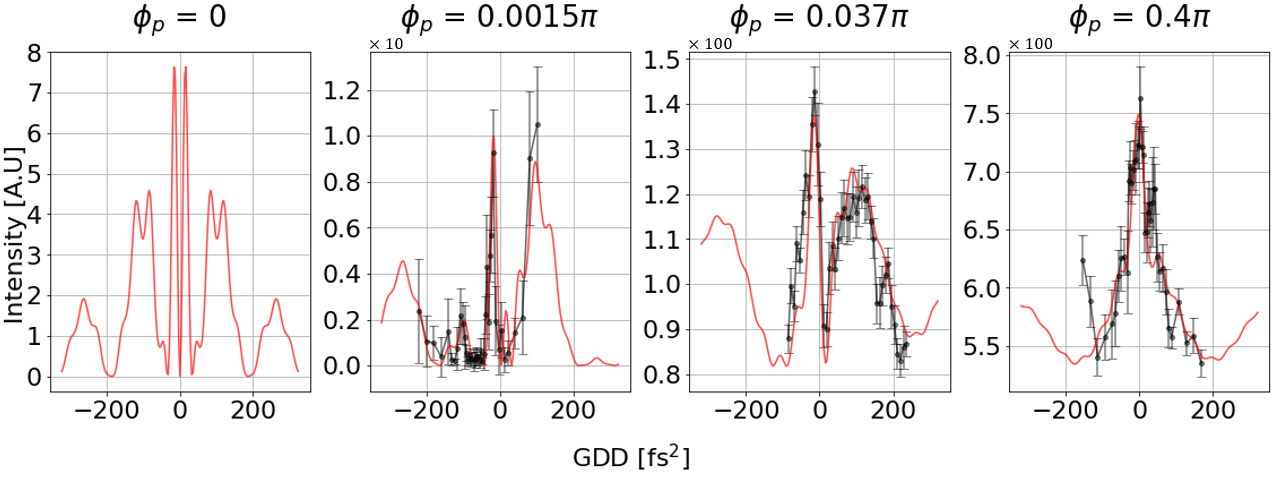}
\caption{{\bf Residue phase sensitivity.} Experimental measurements (black points) and theoretical simulations (red lines) of the SFG signal intensity as a function of the GDD applied to the PDC spectrum by the spectral shaper. Each graph corresponds to a different phase of the pump residue: 0, 0.0015$\pi$, 0.037$\pi$, and 0.4$\pi$. The measured signal captures the time-difference correlation across the entire entangled photon spectrum, where dispersion serves as an analog to time-difference. The strong dependency on dispersion suggests a significant contribution from the spontaneous part of the entangled photons, as the stimulated contribution, which is narrowband, remains unaffected by such variations. This sensitivity is maximized when the phase of the pump residue is near zero, where the contrast of the observed interference patterns is highest.}
\label{dispersion results}
\end{figure}

The high contrast of the measured GDD graph highlights that the contribution of the spontaneous SFG component is dominant (or at least comparable) relative to the stimulated SFG from the single seeded mode (since the two contributions can completely annihilate each other in the SFG signal for some GDD values, e.g. point ``B'' on the graph). In addition, the graphs show an  extreme sensitivity of the SFG signal to the phase of the pump residue, which dramatically affects the form of the GDD scan, its average intensity and its observed contrast. To visualize this phase sensitivity and its effects, Fig.~\ref{dispersion results} shows GDD scans for 4 different phases of the pump-residue: When the residue phase is exactly zero (Fig.~\ref{dispersion results}a, calculation only), the SFG signal is totally nullified at zero GDD, and exhibits a high-contrast and perfectly symmetric scan with a low SFG signal (see Methods for details and derivation). However, this perfect symmetry is fragile and extremely sensitive to the phase of the pump residue, as illustrated in Figs.~\ref{dispersion results}b,c,d, which show GDD-scans for $\phi_p\!=\!0.0015\pi, 0.0365\pi, 0.41\pi$.  In our experiment the residue-phase was stable during the experiment but not absolutely repeatable from one day to another: While the Sagnac interferometer imposed a stable residue phase, the actual phase value was affected by minor misalignment in the Sagnac interferometer that introduced residual asymmetry between the clockwise and counter-clockwise paths in the loop. Since this phase slowly drifted from one day to another, the exact phase of each scan was fitted in the analysis of Fig.~\ref{dispersion results} (the only fitting parameter in our experiment).  

Finally, all the GDD-scans of Fig.~\ref{dispersion results} with various residue phases demonstrate a sharp peak near the zero-GDD point, which is a direct signature of the time correlation of the photon pairs. Specifically, the peak width $\delta\beta\!\approx\!4\pi/\Delta\nu^2\approx 30\,\mathrm{fs}^2$ is inversely proportional to the biphotons bandwidth of $\Delta\nu\!=\! 113$\,THz squared (for $\varphi_r\!=\!0,\pi$ the peak is split). From a physical point of view, this width is also the GDD tolerance for the photon pairs to keep coherence, which reflects their time correlation of the biphotons at $\Delta t\!=\!1/\Delta \nu\!\approx\!9$\,fs. Together with the measured linewidth of $20$Hz for the SFG peak, mentioned above we obtain $ \Delta(\nu_1\!+\!\nu_2)\Delta(t_1\!-\!t_2)\!\approx\!2\!\cdot\!10^{-13}$, showing a giant violation of the classical bound (Eq.~\ref{t-f_entanglement_bound}). The longer-range oscillations in the GDD-scans beyond the central coherent SFG peak are mainly due to the fact that we scan the GDD, rather than the group delay, which indicates that the spectral phase at the center of the spectrum varies quadratically and not linearly across the spectrum.

\section{Discussion}

By combining theory and experiment, we demonstrate that coherent SFG can act as an efficient quantum matched filter, providing a global joint readout of time-difference and frequency-sum correlations. The critical role of time-energy entanglement in the SFG detection is underscored by the fact that our measurements observed only the coherent SFG peak at the pump frequency with no background SFG at other frequencies (down to the dark noise of the RF spectrum analyzer). The inherent coherence of the time-energy entangled photon pairs ensures that the measured SFG spectrum surpasses the shot-noise limit of the pump laser. Specifically, the noise of the pump field amplitude corresponds to one photon in amplitude ($1\sqrt{\text{photons}}/\text{mode}$), whereas the average field amplitude per mode in the experiment was only $\sim\!0.03\sqrt{\text{photons}}/\text{mode}$. Thus, the SFG amplitude of individual mode pairs was undetectable, substantially below this shot-noise limit. And yet, the coherent summation of all the modes across the SPDC spectrum generated a well detected SFG signal, >37dB above the electronic noise floor. 

This highlights that Coherent SFG detection acts as an ideal {\it quantum matched filter} which leverages the ultrafast coincidence nature of SFG and the energy-sum resolution to reject accidental coincidences efficiently in both energy and time, dramatically enhancing the SNR and the detection sensitivity. 

Thus, coherent SFG detection offers a fundamental enhancement over traditional detection schemes, such as SU(1,1) interferometry or coincidence / HOM, since SFG detection is \textit{global,} combining contributions from the entire SPDC spectrum into a single SFG mode, whereas standard methods are \textit{local}, independently measuring specific modes. Our method demonstrates that entanglement detection is most effective through global measurement that probes the entire non-separable system, rather than by isolating individual components. 

This approach of coherent SFG and the dramatic SNR enhancement can be transformative for quantum sensing applications and additional broadband quantum technologies, where the coherent summation opens a door to overcome significant noise limitations. A major example is quantum illumination \cite{PhysRevLett.118.040801, liu2020joint}, which exploits continuous-variables entanglement to enhance the detection of minute signals buried in classical noise, thereby offering substantial advantages for quantum radar systems, secure quantum communication, and advanced imaging techniques. Efficient SFG detection is ideal for realizing the practical benefits of broadband entangled photons in general, and quantum illumination in particular, within noisy and lossy environments.

\section{Methods}

\subsection{Theoretical Model}
We developed a fully-quantum analysis of the three-wave mixing beyond the approximation of an undepleted pump, using a perturbative expansion in powers of the weak fields involved -- the signal-idler pairs in our SFG experiment. While this method is general, and can derive the time evolution of every field up to any desired order, we present here only the aspects of the analysis necessary for modeling our SFG experiment. The complete theory will be published separately. 

The Hamiltonian of multi-mode three-wave mixing between a single pump mode $\hat{a}_p$ and multiple pairs of signal-idler modes $\hat{a}_{\pm\nu}$ is 
$
    \hat{H}=\chi \sum_{\nu} \left(\hat{a}_p\hat{a}_{\nu}^\dagger \hat{a}_{-\nu}^\dagger+\hat{a}_p^\dagger\hat{a}_{\nu} \hat{a}_{-\nu} \right)$, where $\nu$ is the frequency offset of the signal relative the central degenerate mode at $\nu_p/2$ and  $\chi$ the nonlinear coefficient. The Heisenberg equation for the evolution of the pump mode is:
\begin{equation}\label{pump_dynamics_eq}
\dfrac{\mathrm{d}}{\mathrm{d}t}\hat{a}_{p} =\mathrm{i}\left[\hat{H}, \hat{a}_p\right]=-\mathrm{i}\chi\sum_{\nu}\hat{a}_{\nu}\hat{a}_{-\nu}.
\end{equation}
To solve this equation for the pump (and SFG) field, we take a perturbative approach that builds the pump solution as a power series of the weak signal-idler fields. We start with the trivial zeroth-order expression of $\hat{a}_p^{(0)}(t)\!=\!\hat{a}_p\!=\!\alpha_p;\text{  } \hat{a}_{\nu}^{(0)}(t)\!=\!0$, ($\alpha_p$ the complex amplitude of the coherent state of the pump $|\alpha_p\rangle $), which reflects that classically SPDC cannot occur. 

The first order correction is obtained by substituting the zeroth pump solution into the dynamical equations of the signal/idler, which yields the well known solution of parametric amplification under the undepleted-pump approximation:
\begin{equation}\label{standard_OPA_solution}
\hat{a}_{\nu}^{(1)}(t)=\hat{a}_{\nu}\cosh{g}-\mathrm{i}\mathrm{e}^{\mathrm{i}\phi_p}\hat{a}_{-\nu}^\dagger\sinh{g},
\end{equation}
 where $g=|\alpha_p|\chi t$ is the parametric gain with $t=nl/c$ the interaction time within a crystal of length $l$ and refractive index $n$, and $\phi_p$ the pump phase. All operators on the right-hand side are taken at the input to the nonlinear medium. 
 
 To obtain the correction for the pump field due to SFG/depletion, we continue the expansion by substituting the signal/idler fields of the first-order (Eq.~\ref{standard_OPA_solution}) in the dynamical Eq.~\ref{pump_dynamics_eq} of the pump, leading to 
\begin{equation}\label{pump_correction_second_order}
\begin{split}
\hat{a}_{p}^{(2)}(t) = -\mathrm{i}\frac{\sum_\nu\hat{a}_{\nu}\hat{a}_{-\nu}}{2|\alpha_p|}\sinh{2g} - \alpha_p\frac{\hat{N}_s+\hat{N}_i+M}{2{|\alpha_p|}^2}\sinh^2{g}+\alpha_p\frac{\mathrm{i}\hat{H}'}{4{|\alpha_p|}^{3}}(\sinh{2g}-2g),
\end{split}
\end{equation}
where $\hat{N}_s=\sum_\nu\hat{a}_{ \nu}^\dagger\hat{a}_{\nu}$ is the photon number over all the signal modes (and for the idler $\hat{N}_i$ at $-\nu$), $M$ is the number of interacting mode pairs and $\hat{H}'=\hat{H}/\chi=\sum_\nu(\hat{a}_p^\dagger\hat{a}_{\nu}\hat{a}_{-\nu}+\hat{a}_p\hat{a}_{\nu}^\dagger\hat{a}_{-\nu}^\dagger)$ is the unitless Hamiltonian. Eq.~\ref{pump_correction_second_order} is a central result of the SFG analysis. 

Let us pause briefly the derivation to consider the implications of Eq.~\ref{pump_correction_second_order}: First,  within the second order correction, SFG is only due to recombination of twin-modes back to the original pump frequency, whereas ``accidental'' (classical) SFG from uncorrelated modes will appear only in higher order corrections (4th and on). This is an indication of the quantum advantage of the twin-modes SFG over the classical picture.  Second, the number of participating mode pairs $M$ reflects the spectral bandwidth of the PDC light $\Delta\nu$. In an OPO cavity, $M$ is simply the number of cavity comb-teeth within the gain bandwidth, dictated by phase-matching. In our experiment, where the spectrum is continuous,  $M\!=\!\Delta\nu T/2$ reflects the time-bandwidth product, where $\Delta\nu\!\approx\!113$THz and the effective measurement time $T\!=\!1/\delta\nu$ reflects the spectral resolution of our prism-based spectrometer $\delta\nu\!\approx\!0.5$THz, which indicates that the effective number of modes in our experiment was $M\!\approx\!60$. Clearly, the measurement time $T$ can be pushed much longer with a better spectrometer, ultimately limited only by the coherence time (linewidth) of the pump laser, which will push the number of modes to $M\!\approx\!10^{5-6}$ and beyond. And last, note that the only assumption thus far was that the signal-idler fields are weak relative to the pump, but the parametric gain was not limited, allowing $g\geq1$, which indicates that Eq.~\ref{pump_correction_second_order} is applicable also to the high-squeezing multi-photon regime. 

In our experiment however, the parametric gain is low (we estimate $g\approx0.03$), which allows to further approximate Eq.~\ref{pump_correction_second_order} to the second order of $g$, yielding 
\begin{equation}\label{pump_correction_experiment}
\hat{a}_{p}^{(2)}(t) \approx -\mathrm{i}\chi t\sum_{\nu}\hat{a}_{\nu}^{(in)}\hat{a}_{-\nu}^{(in)} - \alpha_p\frac{\hat{N}_{s}^{(in)}+\hat{N}_{i}^{(in)}+M}{2}{(\chi t)}^2,
\end{equation}
where the superscript $^{(in)}$ stands for the intermediate position within the SU(1,1) interferometer (just before the second crystal). Eq.~\ref{pump_correction_experiment} can also be derived by expanding $\hat{a}_p(t)=\mathrm{e}^{\mathrm{i}\hat{H}t}\hat{a}_p(0)\mathrm{e}^{-\mathrm{i}\hat{H}t}$ using the Baker-Campbell-Hausdorff formula, as in \cite{PhysRevA.2023} (but not Eq.~\ref{pump_correction_second_order}).

To express the SFG correction $\hat{a}_{p}^{(2)}$ in terms of the interferometer input, we consider the intermediate field as the output of the first OPA, $\hat{a}_{\nu}-\mathrm{i}\alpha_{p}\chi t\hat{a}_{-\nu}^\dagger$ (Eq.~\ref{standard_OPA_solution} approximated for low gain $g\!=\!\alpha_{p}\chi t$), modified by the spectral phase $\phi_{\nu}$ (from the spectral shaper) and loss $\hat{a}_{\nu}^{(in)}\! =\!\left(\hat{a}_{\nu}\!-\!\mathrm{i}\alpha_{p}\chi t\hat{a}_{-\nu}^\dagger\right)\mathrm{e}^{\mathrm{i}\phi_{\nu}}\sqrt{T}\!+\!b_\nu\sqrt{1\!-\!T}$, where $T$ is the shaper  transmission (assumed uniform across the spectrum) and $b_\nu$ is the field of the loss mode that couples vacuum into the PDC beam. The input states in front of the first crystal is simply $|{\psi_{\pm\nu}}\rangle\! =\! |0\rangle$ for all the spontaneous modes, except for the seeded mode  at $\nu_{sd}$, where $|{\psi_{\nu_{sd}}}\rangle = |\alpha_{sd}\rangle$.

The detected pump field at the output of the Sagnac interferometer mixes the pump+SFG beam (counter-clockwise) from second nonlinear medium with the LO pump (clockwise), which consists of the remaining pump from the first nonlinear medium $\alpha_p$ after slight depletion due to the interaction with the seed and multi-mode vacuum. The LO coherent amplitude is therefore $\alpha_{lo}\!=\!\alpha_p\left(1\!-\!\frac{\left|\alpha_{sd}\right|^2+M}{2}{(\chi t)}^2\right)$, where $N_{sd}\!=\!\left|\alpha_{sd}\right|^2$ is the average number of seed photons at the input. Thus, the Sagnac output field is
\begin{equation}\label{pump_output_field}
\hat{a}_p^{(out)}=\alpha_{r}-\alpha_p\frac{\left|\alpha_{sd}\right|^2+M}{2\sqrt{2}}{(\chi t)}^2+\frac{1}{\sqrt{2}}\hat{a}_{p}^{(2)},
\end{equation}
where $\alpha_r$ is the amplitude of the residue pump due to the imperfect alignment of the Sagnac interferometer (before pump depletion), and the factor of $1/\sqrt{2}$ originates from the transmission of the Sagnac beam splitter. 

Our goal is to calculate the number of photons $\left\langle \hat{N}_{p}\right\rangle = \left\langle \hat{a}_{p}^{\dagger}\hat{a}_{p}\right\rangle$ at the output of the second crystal, considering the effect of lock-in detection, where the phase of all modes is dithered periodically around the locking phase $\phi_\nu(t)\!=\!\phi_\nu\!+\!\phi_d\sin{\omega t}$ (where $\phi_d$ and $\omega$ are the dithering amplitude and frequency), and the SFG intensity is anti-symmetrically integrated in each period to extract $\Delta N_p\!=\!\int dt\left\langle \hat{N}_{p}(t)\right\rangle\sin{\omega t}$. Since the phase dithering operates only on the PDC spectrum, the pump correction $\hat{a}_{p}^{(2)}(t)$ of Eq.~\ref{pump_output_field} can be broken in two: The dithered term 
\begin{equation}\label{dither_amplitude}
    \hat{a}_{dith} = -\frac{\mathrm{i}}{\sqrt{2}}\chi t\sum_{\nu}\hat{a}_{\nu}^{(in)}\hat{a}_{-\nu}^{(in)},
\end{equation}
which varies periodically with the dithering, and the invariant term
\begin{equation}\label{invariant_field}
    \hat{a}_{inv} = \alpha_{r}+\alpha_p\frac{\left|\alpha_{sd}\right|^2-\hat{N}_{s}^{(in)}-\hat{N}_{i}^{(in)}}{2\sqrt{2}}{(\chi t)}^2.
\end{equation} 
Hence, when looking at the number of photons $\left\langle N_{p}\right\rangle = \left\langle a_{p}^{\dagger}a_{p}\right\rangle$ of the output pump we obtain two parts -- quadratic terms and cross terms. The quadratic terms $\hat{a}_{dith}^\dagger\hat{a}_{dith} + \hat{a}_{inv}^\dagger\hat{a}_{inv}$ are phase independent and will cancel out in the demodulation integral, leaving only the cross terms $\hat{a}_{dith}^\dagger\hat{a}_{inv} \!+\! \hat{a}_{inv}^\dagger\hat{a}_{dith}$ to contribute to demodulation. In our experiment $\hat{a}_{inv}\!\approx\!\alpha_r$, since the residual pump amplitude far outweighs the contributions of nonlinear effects in Eq.~\ref{invariant_field}, leading to 
\begin{equation}\label{pump_change_direct}
    {N}_{p}^\mathrm{eff}=\frac{\mathrm{i}}{\sqrt{2}}\chi t\sum_\nu{\left[\alpha_r\hat{a}_{\nu}^{(in)\dagger}\hat{a}_{-\nu}^{(in)\dagger}-\alpha_r^{*}\hat{a}_{\nu}^{(in)}\hat{a}_{-\nu}^{(in)}\right]=\sqrt{2}\chi t\sum_\nu{\text{Im}\left[\alpha_r^{*}\hat{a}_{\nu}^{(in)}\hat{a}_{-\nu}^{(in)}\right]}},
\end{equation}
thereby forming a \textit{homodyne measurement} to extract a single quadrature of the SFG amplitude $\sum_\nu{\hat{a}_{\nu}^{(in)\dagger}\hat{a}_{-\nu}^{(in)\dagger}}$, where $\alpha_r$ acts as the local oscillator. 

\subsection{Experimental Methodology}
A single PPKTP crystal serves the parametric amplifier that both generates the high flux of entangled photons (in forward direction) and measures them via coherent SFG (in the backwards direction) using in a folded configuration of the SU(1,1) interferometer. This crystal is pumped by $\sim\!400$mW power (per direction) from a single-frequency laser at 885\,nm using the Sagnac interferometer described below. This pump wavelength was chosen to maximize the bandwidth of entangled photons by setting the center of the biphotons spectrum near the zero-dispersion wavelength of the PPKTP crystal ($\sim$1790\,nm), where type-0 phase-matching is ultra-broad, ranging from 1.25$\mu$m to >2.5$\mu$m \cite{o2007observation}, exceeding 110\,THz \textendash\;nearly an octave.

To generate an appreciable SFG signal from the entire SPDC bandwidth, it is crucial to maintain a flat spectral phase for all signal-idler frequency pairs in the ultra-broadband spectrum to ensure that every photon pair contributes constructively to the SFG process. This demands precise control of the group delay/velocity dispersion in our SU(1,1) interferometer, achieved by a prism-based shaper of the spectral-phase in the PDC path. This phase shaper, which consists of a prism pair and a reflective 4f-telescope, enables independent control of both the second and the fourth orders of dispersion (following the concept of \cite{shaked2015octave}). In order to tune the dispersion to exact compensation, we measure the spectral phase in real time by observing the spectrum of the ultra-broadband SPDC at the output of the SU(1,1) interferometer with a home-built, prism-based spectrometer. The contrast of the observed SU(1,1) interference was optimized following the guidelines of \cite{nechushtan2021optimal}.

After compensating the dispersion, we also need to stabilize the global phase of the SPDC spectrum relative to the pump. We use for that purpose a Pound-Drever-Hall (PDH) feedback loop, where the phase modulation of the PDH loop is later also used for the lock-in detection of the SFG signal. Conceptually, one could utilize the Intensity of the entangled biphotons at the output of the SU(1,1) interference to retrieve the phase between the photon-pairs and the pump to lock the interferometer to constructive (PDC) or destructive (SFG) interference. However, the SPDC intensity at the single photon regime is too weak and its SNR is low, leading to an unstable and fragile phase lock. Thus, to enhance the measurement SNR and improve the stability and robustness of the phase lock, we seed the SU(1,1) input with a specific signal frequency (1560\,nm), stimulating the PDC to generate a strong matching idler at $\lambda_{i} = 2035$\,nm, whose intensity is sufficient to serve as a reliable reference for phase locking with a satisfactory SNR. The phase lock is implemented using a fast piezoelectric actuator on one of the mirrors in the spectral shaper \cite{goldovsky2016simple}. 

\begin{figure}[ht!]
\centering\includegraphics[width=12cm]{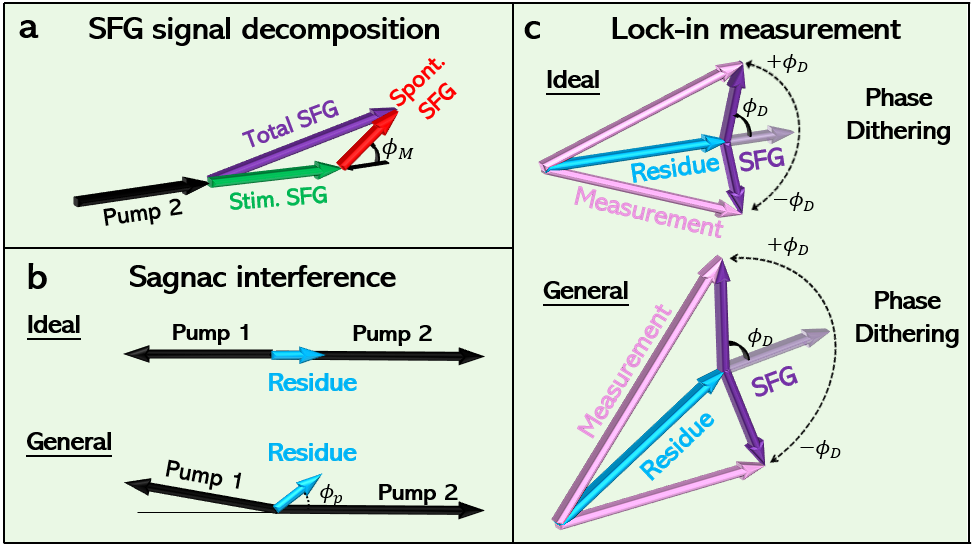}
\caption{{\bf The complex contributions to the Coherent SFG field.} The diagram illustrates the contributions of different phase components in the SFG process, highlighting the roles of the Sagnac residue, spontaneous SFG, and stimulated SFG fields. (a) A diagram showing the decomposition of the total SFG into its spontaneous (red) and stimulated (green) parts, with their phases $\phi_M$ and $\phi_{sd}$, respectively. The black vector represents the pump, which stimulates the second nonlinear interaction. (b) An illustration emphasizing the consequences of misaligned Sagnac, leading to small residue field (blue) with phase $ \varphi_r$. (c) The measured SFG signal (pink) resulting from the interference between these fields, influenced by the phase relationships. Dithering (curved arrow) modulates the phase of the spontaneous SFG and the stimulated SFG relative to the pump, directly influencing the detected signal.}
\label{SFG decomposition}
\end{figure}

Once locked, the phase of the seeded mode is kept by the feedback loop at $\phi_{stim}\!=\!0$  relative to the pump, and the phase of the entire spectrum $\Delta \phi(\nu)$ is stable relative to the seeded mode, since the seeded frequency pair traverses the same optical path as the rest of the SPDC spectrum. By varying the dispersion, the specific phase of each spontaneous frequency-pair $\Delta \phi(\nu)$ can be tuned relative to the single stimulated pair (and the pump), directly influencing the SFG signal, which reflects the coherent sum of the contributions of the spontaneous spectrum and the stimulated mode-pair (see Fig.~\ref{SFG decomposition}a). 

To coherently detect the SFG signal, we employ the Sagnac interferometer (brown section in Fig.~\ref{experimental layout}), where the residual pump field from the first nonlinear interaction (pump 1) serves as a local oscillator for the detection of the SFG contribution from the second nonlinear interaction (pump 2), enabling separation of the SFG contribution from the strong pump background. In the Sagnac interferometer that surrounds the PPKTP crystal, pump 1 and pump 2 propagate in opposite directions (clockwise/counter-clockwise) along the same loop, ensuring that both acquire the same phase. This symmetry leads to robust and stable destructive cancellation at the dark output port of the Sagnac. Consequently, the pump fields cancel each other (almost completely) at the dark port, except for the SFG contribution that breaks the symmetry between the directions (accompanies only pump 2). This destructive interference is crucial for the SFG detection, as it removes most of the intense pump field that would otherwise overwhelm the detector. Note that in reality, a residue pump field still exits from the dark port since the destructive interference is never perfect due to small misalignment in the Sagnac and residual imbalance of the beam-splitter (see Fig.~\ref{SFG decomposition}b). Thus, the detected intensity at the Sagnac dark output (measured with a standard Si photo-detector) is a composite interference of three field contributions: the spontaneous SFG contribution from the wide SPDC spectrum (the measurement target), the stimulated SFG from the pair of seeded modes, and the residue pump field from the Sagnac (see Fig.~\ref{SFG decomposition}a and b). The coherent interference of these fields govern the final results of the measurement and their corresponding phases are the key parameters in their analysis.

Finally, to isolate the desired SFG signal from the much stronger pump residue, we employ lock-in detection by dithering (modulating) the phase of the entire PDC spectrum (spontaneous and stimulated) according to $\phi(t)\!=\!\phi_D\sin\omega t$ using the PZT mirror in the spectral shaper (at frequency $\nu_{D}\!=\!2\pi\!\times\!173\,\mathrm{kHz}$). Due to the coherent nature of the SFG process, this phase dithering is imprinted onto the SFG result, which then interferes with the pump-residue field, modulating the detected SFG intensity, as illustrated in Fig.~\ref{SFG decomposition}c and analyzed theoretically later on. By measuring the spectral intensity of the detected SFG signal at the modulation frequency $I(\nu)$, we effectively isolate the modulated SFG contribution from the large and noisy DC background of the pump-residue. 

Note that this lock-in detection of the SFG signal is simply an optical homodyne measurement of the imaginary quadrature of the SFG amplitude relative to the pump residue (see Fig.~\ref{SFG decomposition}c). For example, in the ideal case of maximum SFG intensity, when the phase of the pump residue is exactly zero and the group delay dispersion (GDD) is perfectly compensated, the lock-in result will actually be nullified (not maximized) since the SFG amplitude is purely real relative to the pump residue, as shown in Fig.~\ref{SFG decomposition}c (top). Intuitively, the perfect symmetry of the phase dithering in the complex plane cancels the intensity difference in the lock-in detection. A phase shift of either the pump residue (due to slight misalignment of the Sagnac) or the SFG due to GDD will break this symmetry in the complex plane and lead to a lock-in detection signal, as shown in  Fig.~\ref{SFG decomposition}c (bottom).

\subsection{Theoretical Calculation of Seeded Coherent SFG}

In order to quantitatively model the SFG signal with a coherent seed, as we have in the experiment, let us consider the contribution of the one seeded mode to the SFG signal in Eq.~\ref{pump_change_direct} separately from the other spontaneous modes.  Substituting Eq.~\ref{dither_amplitude} into the demodulation integral $\left\langle\Delta N_p\right\rangle\!=\!\int dt\left\langle N_{p}(t)\right\rangle\sin{\nu t}$ we obtain the contribution of a single mode pair at frequency $\nu$
\begin{equation}
    \left\langle{\Delta N}_{\nu}\right\rangle\!=\!\frac{\mathrm{i}}{\sqrt{2}}\chi t\left\langle\alpha_r\hat{a}_{\nu}^{(in)\dagger}\hat{a}_{-\nu}^{(in)\dagger}\!-\!\alpha_r^{*}\hat{a}_{\nu}^{(in)}\hat{a}_{-\nu}^{(in)}\right\rangle\!=\!\sqrt{2}A_dT\left|\alpha_r\alpha_{p}\right|{(\chi t)}^{2}\sin{(\varphi_\nu-\varphi_r)},
\end{equation}
where $A_d$ is an integral constant determined by the dithering amplitude and frequency, $\varphi_{\nu}=\phi_{\nu}+\phi_{-\nu}$ is the total phase of the mode-pair at $\nu$ and $\varphi_r$ is the phase of the pump residue (with respect to the pump in the second crystal). For the seeded mode, this intensity is amplified by the stimulation term $\left|\alpha_{sd}\right|^{2}$ ($\alpha_{sd}$ is the coherent amplitude of the seeding field)
\begin{equation}
    \left\langle{\Delta N}_{sd}\right\rangle=\sqrt{2}A_dT\left|\alpha_r\alpha_{p}\right|\left|\alpha_{sd}\right|^{2}(\chi t)^{2}\sin{(\varphi_{sd}-\varphi_r)},
\end{equation}
where $\varphi_{sd}$ is the total phase of the seeded mode-pair in the second nonlinear medium, which is normally stabilized by the active phase lock in the experiment (to $\varphi_{sd}\!=\!0 \;\mathrm{or}\;\pi$). Summing the SFG of the stimulated and spontaneous modes yields
\begin{equation}
\begin{split}
\left\langle\Delta N_p\right\rangle = \sqrt{2}A_dT\left|\alpha_r\alpha_{p}\right|{(\chi t)}^{2}\left[{\left|\alpha_{sd}\right|}^{2}\sin{(\varphi_{sd}-\varphi_r)}\!+\!\sum_{\nu}\sin{(\varphi_{\nu}\!-\!\varphi_r)}\right]. 
\end{split}\label{effective_SFG_intensity}
\end{equation}
The experiment measured the RF spectrum of the SFG intensity, which is proportional to  $\left\langle\Delta N_p\right\rangle^2$ .

By comparing the two terms in Eq.~\ref{effective_SFG_intensity}, we can explain the high sensitivity of the GDD scans to the  residue-phase. The contribution of stimulated SFG (the former term) depends only on the phase of the residue pump (independent of GDD) and can be considered as the ``background amplitude'' of the GDD scan (Fig.~\ref{dispersion results}); while the SPDC contribution (the latter term) is highly sensitive to the spectral change and approaches zero for large GDD values, so it determines the oscillation patterns in Fig.~\ref{dispersion results}. Specifically, the phase of the residue pump will influence both terms: It will determine the value of the average level of the GDD scan in the former term, and affect the shape of the central peak in the latter. For example, in a fully compensated spectrum ($\varphi_{sd}=\pi$ and $\mathrm{GDD}\!=\!0\,\mathrm{fs}^2$) with zero residual phase $\varphi_{r}=0 \text{ or } \pi$, the dither pattern is perfectly symmetric and the SFG amplitude is purely real in the homodyne, which leads to a total cancellation of the SFG signal with a split symmetric central peak (Fig.~\ref{dispersion results}a). However, this perfect symmetry is very easily broken once the residue phase is slightly shifted, as both the experiment and the model show.

\section{Supplementary Information}

\subsection{The non-classical response to loss}

To gain deeper insight into the detection dynamics of our coherent SFG method and the relations between the interfering SFG contributions --  spontaneous and stimulated, we investigated the response of the SFG signal to SPDC loss, regarding the theoretical prediction of Eq.~\ref{effective_SFG_intensity}. SPDC loss affects both photons of a pair, leading to a unique quantum situation, where although the flux of individual photons is reduced by $T$ the biphoton rate is reduced quadratically, as the probability of both photons surviving the SPDC loss is $T^2$.

\begin{figure}[ht!]
\centering\includegraphics[width=12cm]{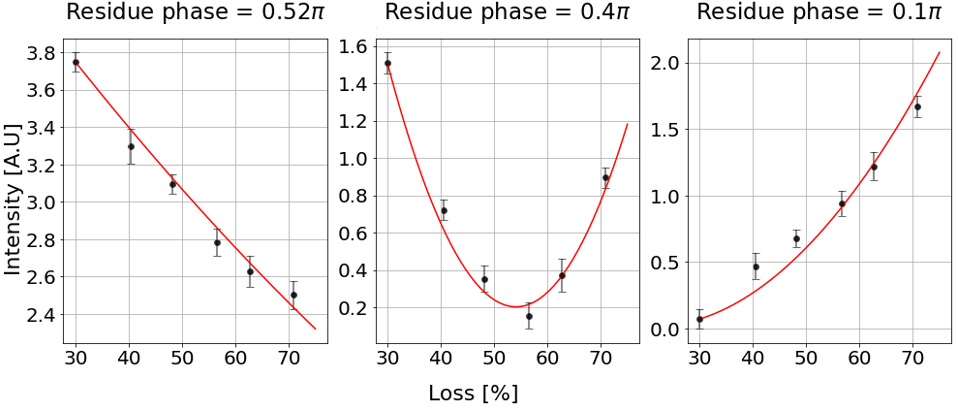}
\caption{{\bf Evidence of non-classicality.} The figure displays the dependency of the measured SFG signal  on losses applied to the spontaneous SPDC spectrum under three different residue phases: 0.5$\pi$, 0.4$\pi$, and 0.1$\pi$. For each scenario, the seed power is adjusted to compensate for its loss, ensuring the classical situation remains unchanged, while altering the balance between the spontaneous and stimulated contributions.}
\label{loss results}
\end{figure}

Figure~\ref{loss results} illustrates three scenarios where  the SPDC losses were applied under different residue phases. In each scenario, the seed power was adjusted to compensate for the loss, ensuring a constant stimulated photon count and isolating the effects on the spontaneous SFG component.  When the spontaneous SFG component is reduced due to SPDC loss, the resultant effect on the SFG signal can be either constructive or destructive, contingent upon the interference between the spontaneous and stimulated components. Specifically, when the stimulated component is relatively small compared with the spontaneous contribution, we may see a flip of trend in the graph (Fig.~\ref{loss results}b). Notably, our theoretical model accurately predicts these interactions, requiring only the residue phase as a fitting parameter.

\begin{backmatter}
\bmsection{Funding}
This project was funded by the Israel Innovation Authority and Elta Systems Ltd. under project No. 70002 and under a Danish-Israeli Eureka project. Funding  is acknowledged also from the European Union’s Horizon Europe research and innovation programme under grant agreement No. 101178170 and from the Israel Science Foundation under grant No. 2208/24. 

\bmsection{Acknowledgments}
We wish to thank Ruben Faibish, David Karasik and Nissan Maskil for helpful discussions. 

\bmsection{Disclosures}
The authors declare no conflicts of interest.

\bmsection{Data Availability Statement}
Data underlying the results presented in this paper may be obtained from the authors upon reasonable request.

\end{backmatter}


\bibliography{bibliography}






\end{document}